# SoftwareX Article

**Title / name of your software**

Elmer FEM-Dakota: A unified open-source computational framework for electromagnetics and data analytics

**Authors / main developers (incl. affiliations, addresses, email)**

Anjali Sandip

Department of Mechanical Engineering, University of North Dakota, Grand Forks, ND 58202, USA, anjali.sandip@und.edu

**Abstract.**

Open-source electromagnetic design software, Elmer FEM, was interfaced with data analytics toolkit, Dakota. Furthermore, the coupled software was validated against a benchmark test. The interface developed provides a unified open-source computational framework for electromagnetics and data analytics. Its key features include uncertainty quantification, surrogate modelling and parameter studies. This framework enables a richer understanding of model predictions to better design electric machines in a time sensitive manner.

*Table 1 – Code metadata (mandatory)*

| Nr | Code metadata description | *Please fill in this column* |
|---|---|---|
| C1 | Current code version | *v2.0* |
| C2 | Permanent link to code/repository used of this code version | https://github.com/AnjaliSandip/ElmerFEM-Dakota |
| C3 | Code Ocean compute capsule | *N/A* |
| C4 | Legal Code License | GPL v3 |
| C5 | Code versioning system used | git |
| C6 | Software code languages, tools, and services used | Elmer FEM (Fortran 2008), Dakota(C++), Bash (C) |
| C7 | Compilation requirements, operating environments & dependencies | Linux |

| C8 | If available Link to developer documentation/manual | -- |
|---|---|---|
| C9 | Support email for questions | anjali.sandip@und.edu |

*1. Motivation and Significance*

Electromagnetic design software coupled with data analytics and parallel programming capabilities can better design electric machines in a time sensitive manner [[1], [2], [3], [4], [5]]. Over the past 5 decades, both, commercial and open-source software with parallel programming capabilities have been developed to design electric machinery [[6], [7], [8], [9]]. A major limitation of the commercial software is that it lacks the ability to customize the source code to better suit the needs of the electromagnetic design process. Open-source software poses a suitable alternative. Traditionally, open-source software lacked tools to perform data analytics which can both provide a richer understanding of model predictions and reduce the computational cost. More recently, there is a growing body of research that has integrated these tools into open-source electromagnetic design software [[10], [11], [12]]. Building on this methodology, the study couples open-source electromagnetic design software, Elmer FEM [13], with data analytics toolkit, Dakota [14]. The overarching goal was to develop an open-source unified framework for electromagnetics and data analytics.

Elmer FEM implements the finite element method (FEM). It is being actively developed by the Finnish IT Center for Science, with a new version released every 12 months. Written in modern Fortran, Elmer FEM's modular structure enables multi-physics simulations. Its parallel programming capabilities coupled with the ever increasing addition of features specific to design of electric machinery has made it, in recent years, an attractive alternative [[10], [15], [16]]. Elmer FEM has a strong user base, over 25,000 downloads/year, which includes academicians and major electric machine design companies such as ABB. The documentation is on par with the development and technical support is provided to users in a timely manner via Elmer user forums [[13], [17]].

Data analytics toolkit, Dakota, is written in C++. It is being actively developed by Sandia National Laboratory, with a new version released every 6 months. The toolkit, built to interface with simulation codes, is supplemented by extensive documentation. One of the primary advantages of Dakota is it enables the user to perform a wide range of data analyses by making minor modifications to its input file. A graphical user interface, parallel computing, and robust post-processing capabilities are some of the other key features of this program [14]. Dakota has a world-wide user community. It has been solicited by industry and major research organizations such as NASA [[18], [19]].

## 2. Software Description

The program integrates the capabilities of electromagnetic design software, Elmer FEM, and data analytics toolbox, Dakota.  It is written in bash and can be run in serial or parallel.  It uses functionality from the underlying packages.  The program was developed to integrate Dakota version 6.10 with Elmer FEM version 8.1.   It can be extended to other versions of Elmer FEM and Dakota with minor modifications.  The interfacing approach is non-intrusive.  The clear distinction between the files associated with Dakota and Elmer FEM in the integrated computational framework enhances its usability. The user is expected to set up the Dakota and Elmer FEM project files for their application. There are no constraints on the choice of the Dakota analysis method or Elmer FEM solver. The program along with examples are provided in the software repository.

Dakota invokes the interface program which provides the necessary instructions to perform data analysis on electromagnetic model/s.  The interface program is divided into three parts: pre-processing, run and post-processing.  In the first part of the program, dprepro, a command line tool, is used to transfer a new set of values for the chosen input parameters from Dakota to Elmer FEM.  The complexity of the input parameter/s could range from boundary conditions to geometry and parameterized mesh [14].  To run the simulation for the new set of input parameter values, ElmerSolver is invoked using Dakota's external simulation interface.  Once the simulation has been completed, quantities of interest are extracted from the simulation results and transferred to the Dakota results file.  If the stopping criteria are met, then the analysis is terminated and the results of the data analysis are recorded, else the control is transferred back to the pre-processing section of the interface program and the process repeats itself.  Communication between Elmer FEM and Dakota is illustrated in Fig. 1.

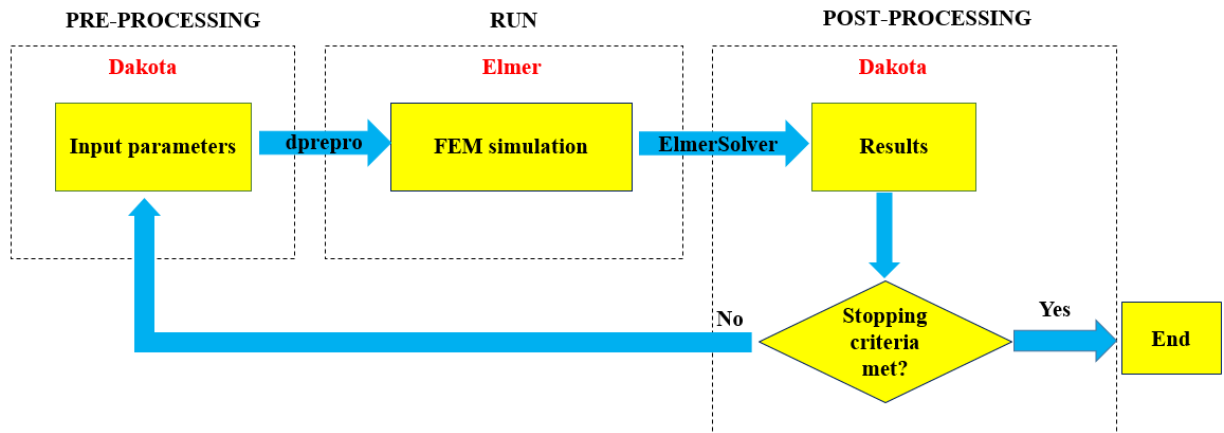

Fig. 1.  Flowchart of the Elmer FEM – Dakota interface program

## 3. Illustrative Examples

The interface program was validated against a 3D, steady state benchmark test: magnetic field induced by a potential difference applied over a wire of finite length [20, 21]. The material properties of the wire are that of copper. The wire is aligned in the z direction and is surrounded by air. The objective was to determine the effect of changes in wire radius (+ 2 m) and electric conductivity (+ 2 x 10⁶ A/m-V) on maximum field value for joule heating.

Analytical model was implemented in MATLAB. The numerical model was built in Elmer FEM using unstructured, hexahedral mesh elements and quadratic linear elements. Elmer FEM's electromagnetic solver, WhitneyAV, was chosen for this study. The solver implements classic Lagrange interpolation and edge element basis functions to approximate the scalar and the vector potentials, respectively [21].

Latin hypercube sampling study was performed, with a sample size of 400. The input parameters were distributed normally. Asynchronous local parallelism was implemented, 4 evaluations performed concurrently. The results from the numerical model were in good agreement with the analytical model as shown in Fig. 2. The numerical model predicts that for the changes in the wire's material and geometric parameters, there is a 95% probability that the maximum field value for joule heating would vary between 0.004 and 0.011 watts. In addition, design and analysis of computer experiments (DACE) technique was employed to generate training data, a sample size of 50 was chosen. A linear regression fit was applied to the training data to construct the global surrogate model. Latin hypercube sampling study was then performed on the global surrogate model. The DACE technique was shown to reduce the total run time by 87% for the benchmark test while maintaining the predictive accuracy of the numerical model.

The benchmark test demonstrates the key capabilities of the Elmer FEM-Dakota interface program: uncertainty quantification, surrogate modelling and parameter studies. The files required to reproduce the benchmark test are provided in the software repository.

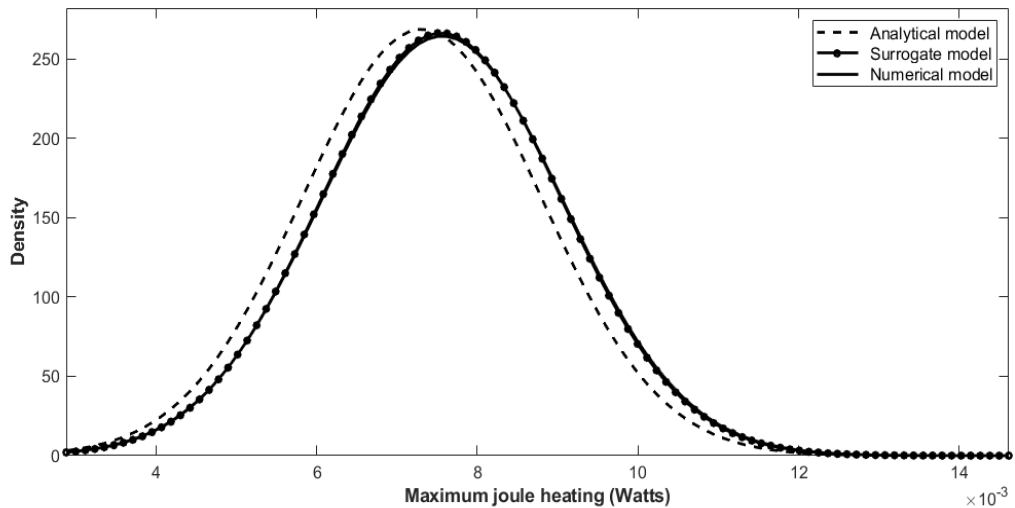

Fig. 2. Comparison between analytical, numerical and surrogate model predictions for maximum joule heating. The probability density function histograms obtained from the results of the latin hypercube sampling study were fit to a normal distribution. There was a 4% discrepancy in the central tendency predictions between the numerical and the analytical model.

4. *Impact and conclusions*

Elmer FEM-Dakota interface program was developed to provide a unified open-source computational framework for electromagnetics and data analytics. This unification enables a richer understanding of model predictions in a time sensitive manner. It can be run in serial or parallel. The following features enhances the framework's usability: (1) There are no constraints on the choice of the Dakota analysis method or Elmer FEM solver, (2) The ability to perform a wide range of data analyses by making minor modifications to the Dakota input file, and (3) The clear distinction between the files associated with Dakota and Elmer FEM in the integrated computational framework.

In this study, the interface program was validated against a benchmark test in electromagnetics: magnetic field induced by a potential difference applied over a wire. The objective was to quantify the effect of uncertainties in geometric and material parameters on joule heating. Numerical simulations, developed in Elmer FEM, were run in parallel. The results were in good agreement with the analytical model as shown in Figure 2. Furthermore, surrogate-based global minimization was performed. This reduced the total run time by 87%.

The framework developed in this study provides tools to better design electric machines at reduced computational cost. For example, optimizing the design of electric machinery can be computationally expensive as it requires a full numerical solution of the electromagnetic field for each iteration. Employing the framework's surrogate modelling capabilities can lead to significant reduction in the computational cost. The framework can also be applied to quantify the uncertainty in critical geometric parameters, such as air gaps, leading to improved predictions [22].

The interface program's primary application area is electromagnetics, but it can be easily extended to other multi-physics applications, such as thermo-mechanics, fluid-structure interactions, etc.

*Role of the funding source*
This research did not receive any specific grant from funding agencies in the public, commercial, or not-for-profit sectors.

*Acknowledgements*
The author acknowledges the support provided on the Dakota and the Elmer user forums.